\documentclass[reqno]{amsart}
\usepackage{graphicx,amsmath}
\usepackage{float}
\usepackage[utf8]{inputenc}
\usepackage{color}
\usepackage{cite}
\usepackage{fancyhdr}
\usepackage[margin=2cm]{geometry}
\usepackage{appendix}
\usepackage{multirow}
{ 
\sloppy
\begin{document}
\title{The role of the initial states for non-Fourier heat equations}

\author{R. Kovács$^{123}$}

\address{
$^1$Department of Energy Engineering, Faculty of Mechanical Engineering, BME, Budapest, Hungary
$^2$Department of Theoretical Physics, Wigner Research Centre for Physics,
Institute for Particle and Nuclear Physics, Budapest, Hungary
$^3$Montavid Thermodynamic Research Group
}

\date{\today}

\begin{abstract}
There are several models for heat conduction - non-Fourier equations - in the literature that are important for various practical problems. These models manifest themselves in partial differential equations, and the application of which requires developing efficient and reliable solution methods. In the present paper, we focus on the analytical solutions of two non-Fourier models, specifically on the Maxwell-Cattaneo-Vernotte and Guyer-Krumhansl equations, as they share an established thermodynamic background, and find numerous applications. Although initial conditions are usually homogeneous in space in many situations, real applications can easily point beyond such simple initial state. Therefore, we aim to investigate the consequences of nonhomogeneous initial conditions, emphasising the physical requirements to keep the solution physically admissible. We conclude the calculations with a method for determining the initial time derivatives in consistence with thermodynamics, avoiding contradictions.

\end{abstract}
\maketitle

\section{Introduction}
In physics and engineering, models are used as mathematical images of the world we experience around ourselves. Therefore, the actual task defines which elements we include in the model and which phenomena we want to understand, predict, and characterise quantitatively. Throughout this paper, we focus on heat conduction as the primary phenomenon. 

Fourier's well-known model, i.e., when the heat flux ($\mathbf q$) is being proportional with the temperature gradient ($\nabla T$),
\begin{align}
\mathbf q = - \lambda \nabla T \label{fourier}
\end{align}
with $\lambda$ being the thermal conductivity, leads to the simplest heat conduction equation that satisfies the II. law of thermodynamics,
\begin{align}
\partial_t T = \alpha \Delta T, \label{fouuu}
\end{align}
where $\alpha=\lambda/(\rho c)$ is the thermal diffusivity with mass density $\rho$ and specific heat $c$, both assumed constant here; $\partial_t$ denotes the partial derivative respect to time $t$, and $\Delta$ is the Laplacian. Although it is applicable for most of the engineering tasks, in recent decades, several situations have been discovered for which Fourier's law loses its validity. Such cases can be a low-temperature environment \cite{Tisza38, Pesh44, McNEta70a}, nanostructures \cite{Zhang07b, LebEtal11, JouCimm16}, heterogeneous materials \cite{Sobolev94, Botetal16, Sobolev16, FehEtal21, FehKov21} and even low-pressure states for fluids \cite{MulRug98, Struc05, StrTah11, RugSug15}. These problems can be modelled with a non-Fourier equation, in which Eq.~\eqref{fourier} is replaced with a (partial) differential equation, and, correspondingly, \eqref{fouuu} gets generalized to a partial differential equation that is higher-order in time. The level of extension depends on the particular situation and the chosen approach, hence, several models exist in the literature. 

It seems inevitable to investigate and understand the mathematical properties of non-Fourier models, as one of these may be a new standard model in the future, substituting the Fourier heat equation. In the present study, we want to investigate two of them, the Maxwell-Cattaneo-Vernotte (MCV) \cite{Cattaneo58, Vernotte58} and the Guyer-Krumhansl (GK) \cite{GuyKru66a1} equations, as they are the simplest, thermodynamically compatible extensions of Fourier's law. Recently, both numerical and analytical solution methods have been developed for these models with special attention to the boundary conditions \cite{Kov18gk} in order to avoid unphysical results such as negative absolute temperature \cite{Zhukov16, Zhu16b}. Now, we turn our attention to the initial conditions. Typically, the models are solved with steady initial states and homogeneous field variables. When this is not the case, the situation may be surprisingly more difficult to overcome, and one must take the first step with care to avoid misleading assumptions. 
Consequently, here we consider a situation in which the initial temperature distribution is nonhomogeneous, thus having a non-equilibrium initial state. For the Fourier heat equation \eqref{fouuu}, that problem is almost trivial, as it only requires knowledge the temperature profile at the initial time instant. However, for a non-Fourier equation, the initial time deriatives are also required, which is restricted by a  more complicated constitutive equation. Now, such constitutive relationship is a differential equation, needing further considerations about the initial conditions. This paper aims to answer this question through analytical solutions of the aforementioned heat conduction
models, which reflect the essential physical aspects. 
In what follows, we first present the concept of the Fourier heat equation, forming the basis of the analytical solution technique  also used for non-Fourier models. After that, 
 we can move on to the more complex aspects of non-Fourier models and reveal how the initial temperature state influence the initial time derivatives. We highlight the critical steps by analytically solving the MCV and GK equations. 

\section{Demonstration using Fourier's law}
\begin{figure}[H]
\centering
\includegraphics[width=12cm,height=6cm]{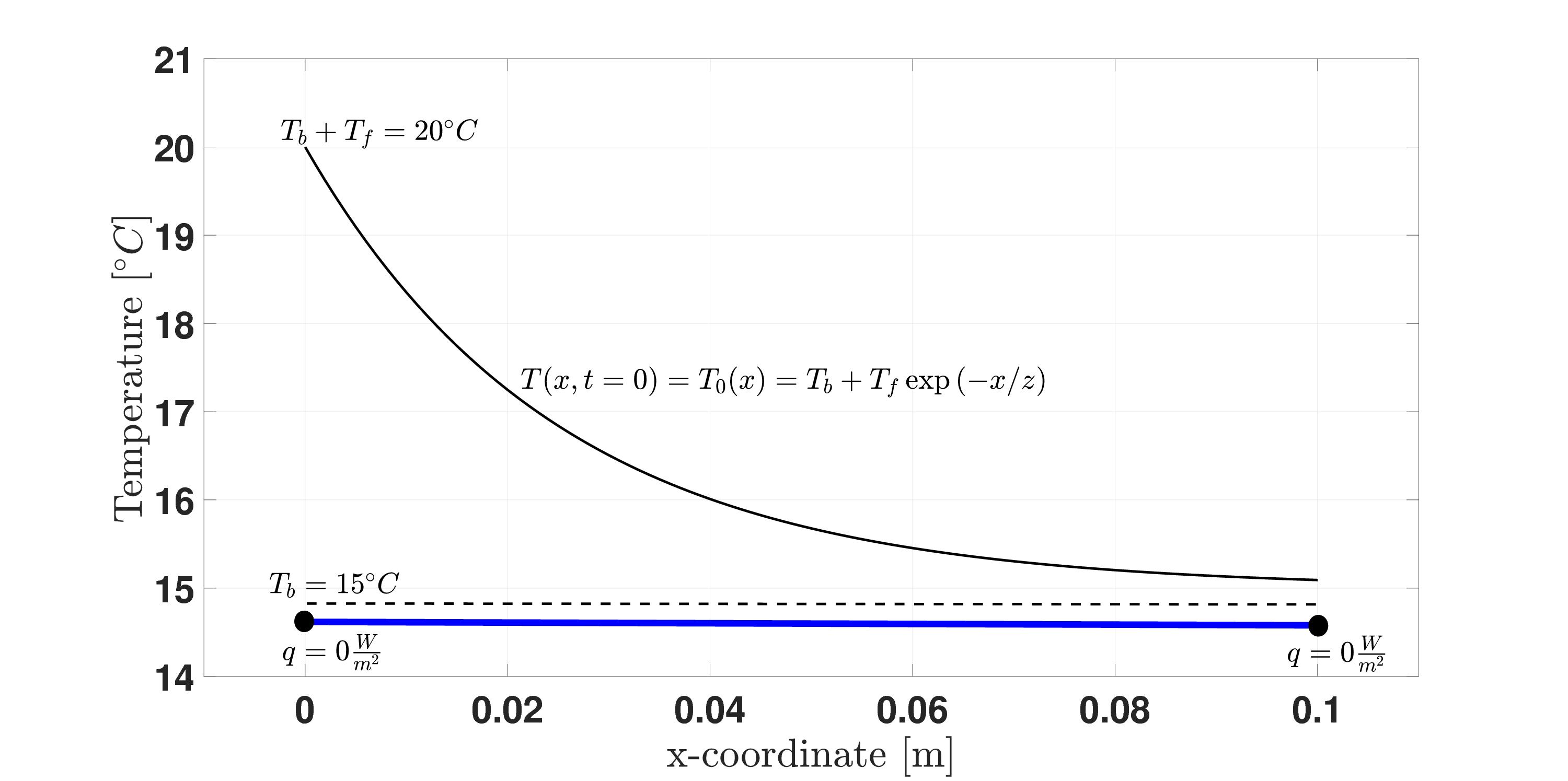}
\caption{The schematic presentation of the initial and boundary conditions, with $z=0.025$ m, $T_\textrm{b}=15$ $^\circ$C, $T_f=5$ $^\circ$C, $L=0.1$ m.}
\label{fig1}
\end{figure}

\noindent In one space dimension, Fourier's law \eqref{fourier} reads,
\begin{align}
q=-\lambda \partial_x T, \label{fourier2}
\end{align}
which is a constitutive equation and becomes mathematically and physically complete together with the balance equation of internal energy ($e=cT$),
\begin{align}
\rho c \partial_t T + \partial_x q = 0, \label{ebal}
\end{align}
for rigid conductors without heat sources. The simplicity of Eq.~\eqref{fourier2} suggests to use the temperature $T$ as the only field variable, i.e., substituting \eqref{fourier2} into \eqref{ebal} results in
\begin{align}
\partial_t T = \alpha \partial_{xx} T. \label{FT}
\end{align}
Despite the fact that Eq.~\eqref{FT} would be adequate for certain investigations, we do not intend to eliminate any of the variables as the initial pair of equations \eqref{fourier2}--\eqref{ebal} is more suitable for our
present purposes. Let us consider the following initial and boundary conditions,
\begin{align}
q(x=0,t)=q(x=L,t)=0 \frac{\mathrm{W}}{\mathrm{m}^2}; \quad T(x,t=0)=T_0(x)=T_\textrm{b} + T_f \exp{(-x/z)}, \label{icbc}
\end{align}
which represent a rod of length $L$ with adiabatic ends. That set of initial and boundary conditions \eqref{icbc} prescribes a non-equilibrium situation as an initial state. Regarding the initial temperature distribution, we choose an exponential decay, illustrated in Figure \ref{fig1}. 

The practical importance of such an initial temperature distribution is that radiation (incident from the left) absorbed by the body induces such a temperature profile. For instance, this is realistic in a heat pulse experiment in which some part of the flash energy is absorbed in a semi-transparent body. Accordingly, $z$ represents a penetration depth quantity, while $T_f$ is the temperature rise at the front side, thus $T_b + T_f$ together form the front side temperature as $T_b$ is the temperature of the body before the flash occurs (e.g., ambient temperature). 
We expect the nonhomogeneous temperature profile to equilibrate in time, regardless of the applied heat conduction model. 

\begin{figure}[H]
\centering
\includegraphics[width=17cm,height=6.7cm]{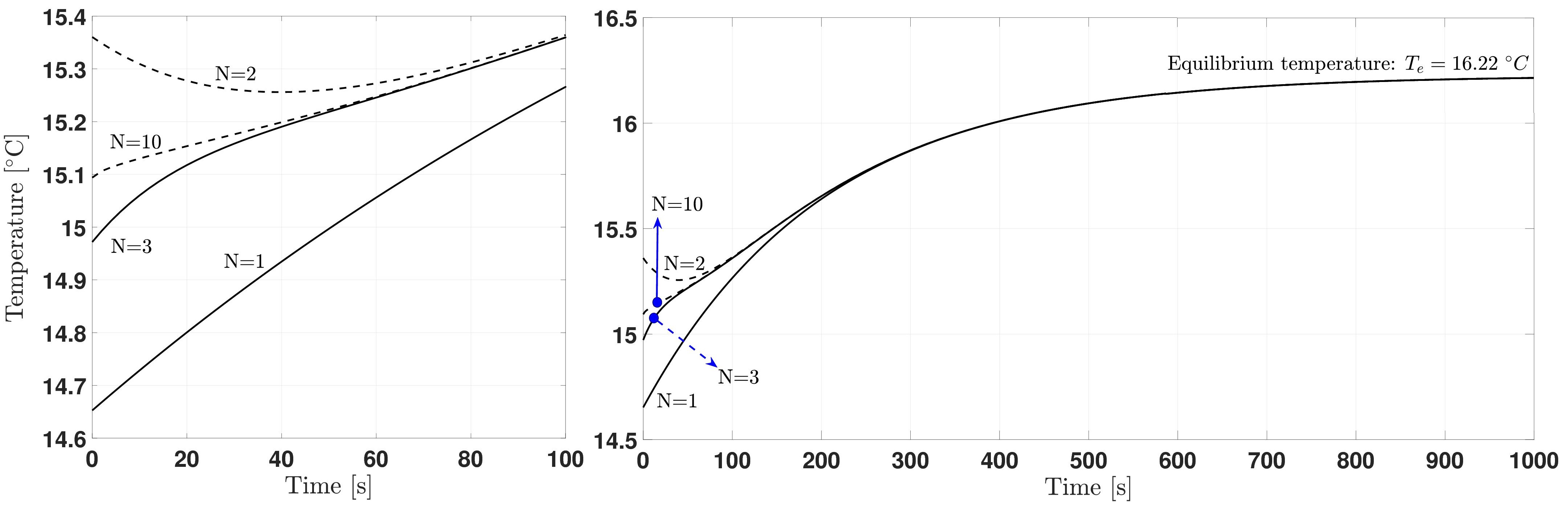}
\caption{Demonstrating the convergence of \eqref{thist}, rear side ($x=L=0.1$ m) temperature history. The left side figure magnifies the initial time interval.}
\label{fig2}
\end{figure}

\begin{figure}[]
\centering
\includegraphics[width=14cm,height=7.5cm]{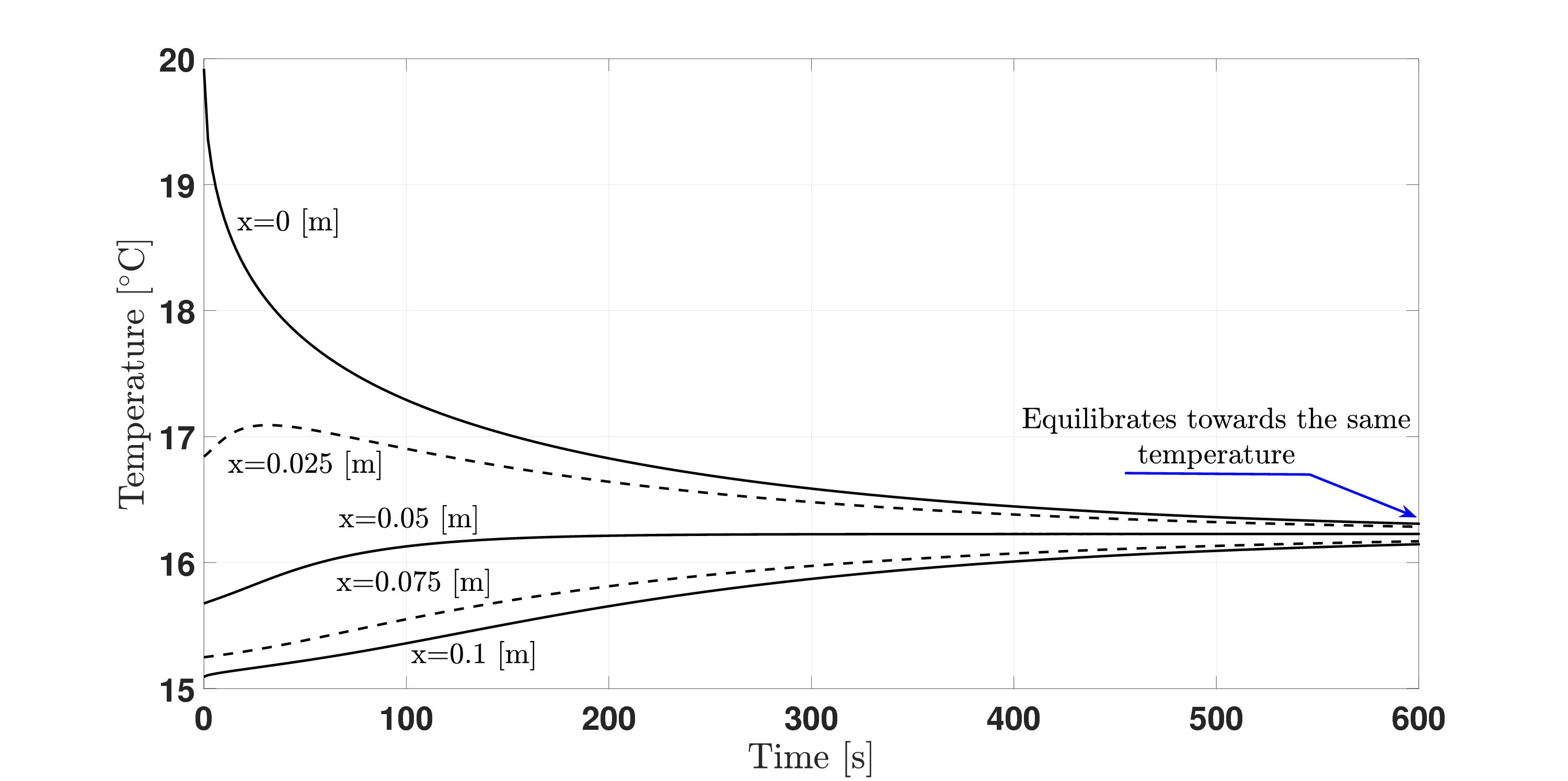}
\caption{Temperature history for different points ($x=\{0, 0.025, 0.05, 0.075, 0.1\}$ m), using $N=50$ terms.}
\label{fig3}
\end{figure}

As learned from numerical solutions for Fourier and beyond-Fourier
heat conduction problems \cite{RietEtal18}, it is advantageous to compute both $T$ and $q$. Here, we determine them analytically, using Galerkin's method.
However, in order to remain consistent with the non-Fourier models and highlight the critical steps, we choose a different approach and solve the model analytically using Galerkin's method. Let us suppose that 
\begin{align}
q(x,t)=\sum_{j=0}^N a_j(t) \phi_j(x), \quad T(x,t)=T_\textrm{b}+\sum_{j=0}^N b_j(t) \varphi_j(x) \label{g1}
\end{align}
where $\phi_j(x)$ and $\varphi_j(x)$ should be compatible with each other due to the spatial derivatives that appear in Eqs.~\eqref{fourier2} and \eqref{ebal}. Suitable functions are found earlier in \cite{FehKov21, Kov18gk}, that is, $\phi_j(x)=\sin(j \pi x/L)$ and $\varphi_j(x)=\cos(j \pi x/L)$. Here, the modes $\phi_j$ fulfil the boundary conditions and form a
complete orthogonal set of functions among the square integrable
functions, and the functions $\varphi_j$ also provide a complete
orthogonal set \cite{Fulop07}. With the help of them, we can transform the partial differential equation to a set of ordinary differential equations. After substituting Eq.~\eqref{g1} into \eqref{fourier2} and \eqref{ebal}, we obtain
\begin{align}
a_j =&\lambda \frac{j \pi}{L} b_j, \label{g2f}\\
\rho c \frac{\textrm{d} b_j}{\textrm{d} t}=& -\frac{j \pi}{L} a_j.\label{g2e}
\end{align}
Furthermore, $a_0=0$, and the cosine series expansion of $T_0(x)$ serves as the initial condition:
\begin{align}
b_0 =& \frac{1}{L} \int\displaylimits_0^L \big(T_0(x)-T_\textrm{b}\big) \textrm{d}x = \frac{T_f z}{L} \Big ( 1 - \exp(-L/z) \Big),  \\ 
b_j (t=0) = b_{j0}=& \frac{2}{L} \int\displaylimits_0^L \big(T_0(x)-T_\textrm{b}\big) \cos\left(\frac{j \pi x}{L} \right) \textrm{d}x = \frac{2 T_f z L e^{-L/z}}{L^2 + j^2 \pi^2 z^2} \left ( e^{L/z} - (-1)^j \right ). \label{f1}
\end{align}
Finally, we found the solution to be
\begin{align}
T(x,t)=T_\textrm{b}+{b_0} + \sum_{j=1}^N b_{j0} e^{-\alpha_j t} \cos\left(\frac{j \pi x}{L} \right), \quad \textrm{with} \quad \alpha_j = \frac{\lambda}{\rho c} \frac{j^2 \pi^2 }{L^2}, \label{thist}
\end{align}
for which Figure \ref{fig2} shows the convergence using the set of the following parameters: $\rho=2000$ kg/m$^3$, $c=500$ J/(kg$\cdot $K), $\lambda=5$ W/(m$\cdot $K), with the initial conditions of Figure \ref{fig1}. Figure \ref{fig3} shows the temperature for various spatial points. While the results followed from a straightforward calculation and seemed basic, it makes the fundamental aspects of non-Fourier models apparent in the next section.

\section{Initial states vs.~non-Fourier equations}
We start this section with the MCV equation, the first hyperbolic extension of the Fourier heat equation. The MCV model is - unfortunately - restricted to low-temperature heat conduction problems \cite{Naretal75}, and thus, it has less practical relevance in engineering, as it has been concluded by Auriault \cite{Aur16}, in agreement with our experience \cite{SudEtal21}. However, it serves as an excellent example of how careful we must be when using an unconventional model. Subsequently, this impression is strengthened further by investigating the GK equation for which neither the initial conditions nor the boundary conditions work as in the Fourier case \cite{FehKov21}.
\subsection*{Maxwell-Cattaneo-Vernotte equation}
The MCV model extends the Fourier equation by the time derivative of the heat flux, i.e.,
\begin{align}
\tau \partial_t q + q = - \lambda \partial_x T, \label{mcv0}
\end{align}
in which a new parameter $\tau$, called relaxation time, appears. As a direct consequence of the II.~law of thermodynamics, $\tau$ and $\lambda$ are not independent of each other due to the Onsagerian relations, which property becomes crucial for nonlinearities, e.g., for temperature-dependent parameters \cite{KovRog20}. 

Analogously to the Fourier model, it is possible to eliminate one of the variables ($q$ or $T$, using Eq.~\eqref{ebal}), resulting in
\begin{align}
\tau \partial_{tt} T + \partial_t T &= \alpha \partial_{xx} T, \quad \textrm{or}  \label{mcvt}\\
\tau \partial_{tt} q + \partial_t q &= \alpha \partial_{xx} q, \label{mcvq}
\end{align}
depending on which variable we choose in a linear situation. While $q$ is not conventional to use, it could be helpful in certain cases \cite{FehKov21}. Nevertheless, in general, we recommend not eliminating any of the variables. As it is visible, the set of initial conditions seemingly depends on our decision, which path we choose. For Eq.~\eqref{mcvq}, one should define the heat flux and its time derivative at the initial time instant, however, none of them is directly measurable. In parallel, regarding Eq.~\eqref{mcvt}, we have the initial temperature distribution $T_0(x)$, but what can we tell about its time derivative? If one follows the papers of Moosaie \cite{Moosaie07, Moosaie08}, or Tung and Fong \cite{TungFong11}, one could conclude that it is possible to safely assume the time derivative to be zero since the thermodynamic origin of the $T$-representation is entirely hidden in their approach.  As it turns out soon, this would be a seriously misleading assumption.

Applying again the ansatz of \eqref{g1} on \eqref{mcv0}, we obtain
\begin{align}
\tau \frac{\textrm{d} a_j}{\textrm{d} t} + a_j = \lambda \frac{j \pi}{L} b_j,
\end{align}
together with \eqref{g2e}. Let us eliminate the variables again, i.e., eliminating the heat flux ($q$), we have
\begin{align}
\textrm{$T$-representation:}& \quad \quad \tau \frac{\textrm{d}^2 b_j}{\textrm{d} t^2} + \frac{\textrm{d} b_j}{\textrm{d} t} + \alpha_j b_j = 0, \quad \quad \alpha_j= \frac{\lambda}{\rho c} \frac{j^2 \pi^2 }{L^2} \\
\textrm{initial conditions:}& \quad \quad b_j(t=0)=b_{j0}, \quad \textrm{and} \quad \frac{\textrm{d} b_j}{\textrm{d} t}\bigg\rvert_{t=0} = -\frac{j \pi}{\rho c L} a_j(t=0), \nonumber
\end{align}
in which the time derivative of $b_j$ follows from the balance of internal energy \eqref{ebal} in case of no heat sources, otherwise the heat source would also appear here. Thus, the MCV equation can be solved if the heat flux is known at the initial time instant, directly \textit{determining} the initial time derivative of temperature, which is not visible from \eqref{mcvt}. Now, there is an advantageous property of most non-Fourier heat equations: their steady-state coincides with the Fourier equation. Consequently, if and only if the initial state is close to steady-state, then the heat flux can be calculated easily using \eqref{fourier2}, and by having $a_j(t=0)$, the model can be solved. For non-zero $T_0(x)-T_\textrm{b}$, it means a non-zero time derivative of temperature, too. In parallel, if that $T_0(x)$ represents a state far from equilibrium, then we cannot use \eqref{fourier2} anymore for such a purpose. Hence the way of how to determine the initial time derivative of temperature remains an open question. We want to recall the fact that, for our application-motivated present set of initial and boundary conditions \eqref{icbc}, the initial state is not a steady-state.

In the case when $q$ is chosed to be the primary field variable, we obtain
\begin{align}
\textrm{$q$-representation:}& \quad \quad \tau \frac{\textrm{d}^2 a_j}{\textrm{d} t^2} + \frac{\textrm{d} a_j}{\textrm{d} t} + \alpha_j a_j = 0, \quad \quad \alpha_j= \frac{\lambda}{\rho c} \frac{j^2 \pi^2 }{L^2} \\
\textrm{initial conditions:}& \quad \quad a_j(t=0)=a_{j0}, \quad \textrm{and} \quad \frac{\textrm{d} a_j}{\textrm{d} t}\bigg\rvert_{t=0} = \frac{1}{\tau}\Big(\lambda\frac{j \pi}{ L} b_{j0} - a_{j0} \Big ), \nonumber
\end{align}
which reveals that the initial time derivative of heat flux is zero if and only if the Fourier law is applied to determine the initial heat flux due to \eqref{g2f}. In both representations, \eqref{f1} is used to determine the solutions, plotted in Figure \ref{fig4} for demonstration.

A more complicated way to find the proper initial condition is if one substitutes the function $T_0(x)$ into the constitutive equation \eqref{mcv0}, as it is usual with the Fourier equation,
\begin{align}
q(x,t=0) = - \lambda \partial_x T_0(x) + C(x), \label{mcvic0}
\end{align}
where $C(x)$ appears by solving the differential equation \eqref{mcv0} for $q$, and taking $t=0$. This is reasonable since \eqref{mcv0} describes the material, with which the initial condition must be compatible. Moreover, the balance equation \eqref{ebal} leads to
\begin{align}
\partial_t T = - \frac{1}{\rho c} \partial_x q = \frac{1}{\rho c} \Big ( \lambda \partial_{xx} T_0(x) - \partial_x C(x) \Big ).
\end{align}
From that point of view, $C(x)$ is an uncertainty that must be restricted somehow. Exploiting the knowledge that steady-states are identical, one would think that $C(x)$ should be zero. However, this would be misleading: the boundary conditions do have a role at this step. In other words, $q(x=0,t)=-\lambda \partial_x T_0(x=0) + C(x=0)=0$ and the same for $x=L$ holds, therefore $C(x)$ depends on the initial temperature distribution and the boundary conditions, too. Moreover, as it could depend on $x$, one has the freedom to choose a suitable function for dynamic situations. However, if $C(x)$ is zero everywhere except on the boundary, then $C(x)$ is not necessarily differentiable at $x=\{0,L\}$, or $\partial_x C(x)$ could be difficult to find and it definitely is not zero everywhere. While it seemed reasonable, finding a proper initial time derivative is not straightforward.

Instead of sticking to the conventional approach, Galerkin's method can offer deeper insight, even with nonzero (and time-dependent) boundary conditions.. In that case, it is possible to separate the boundaries \cite{FehKov21} and deal with an adiabatic problem. Then, the basis functions ($\phi$ and $\varphi$) automatically satisfy the boundary conditions and provide a more explicit and straightforward approach. 

Consequently, realising of these aspects was possible due to the chosen solution method. Galerkin's method revealed which modes must be used to obtain a physically compatible set of initial conditions. However, without any prior knowledge of the non-Fourier heat conduction constitutive law, it would be difficult to do so. If one starts with utilising only the $T$-representation of the heat equation, the compatibility between the initial conditions and the heat conduction constitutive relationship can be easily violated. 

\subsection*{Guyer-Krumhansl equation}
While the GK equation was incredibly helpful in modelling low-temperature phenomenon of second sound in solids \cite{GK66}, it turned out recently that it is also applicable for macroscale heterogeneous materials at room temperature, when over-diffusion appears, using a continuum background \cite{VanFul12, JozsKov20b}. Therefore, this model may have greater practical importance later, and we feel it necessary to investigate the characteristics of the GK equation. 
It introduces a new spatial derivative term into the constitutive equation, compared to \eqref{mcv0},
\begin{align}
\tau \partial_t q + q = - \lambda \partial_x T + l^2 \partial_{xx} q, \label{gk0}
\end{align}
where $l^2$ has originally been interpreted as the squared mean free
path of kinetic theory. However, as we utilise a continuum thermodynamic approach instead of the kinetic theory, the coefficients $\tau, \lambda, l^2\geq0$ are restricted only by the II.~law of thermodynamics, and can be fitted to experiments. That new term makes it more difficult to properly handle the boundary conditions \cite{BallEtal20}. The usual interpretation of boundaries by Fourier is no longer valid. The $T$ and $q$ representations are
\begin{align}
\tau \partial_{tt} T + \partial_t T &= \alpha \partial_{xx} T + l^2 \partial_{txx}  T, \label{gkt}\\
\tau \partial_{tt} q + \partial_t q &= \alpha \partial_{xx} q + l^2 \partial_{txx} q, \label{gkq}
\end{align}
with exploiting \eqref{ebal} again. Having \eqref{gkt}, the prescribed $\partial_x T$ is not a valid boundary condition, and leads to unphysical results \cite{Zhukov16, Zhu16b}.

\begin{figure}[H]
\centering
\includegraphics[width=8cm,height=7cm]{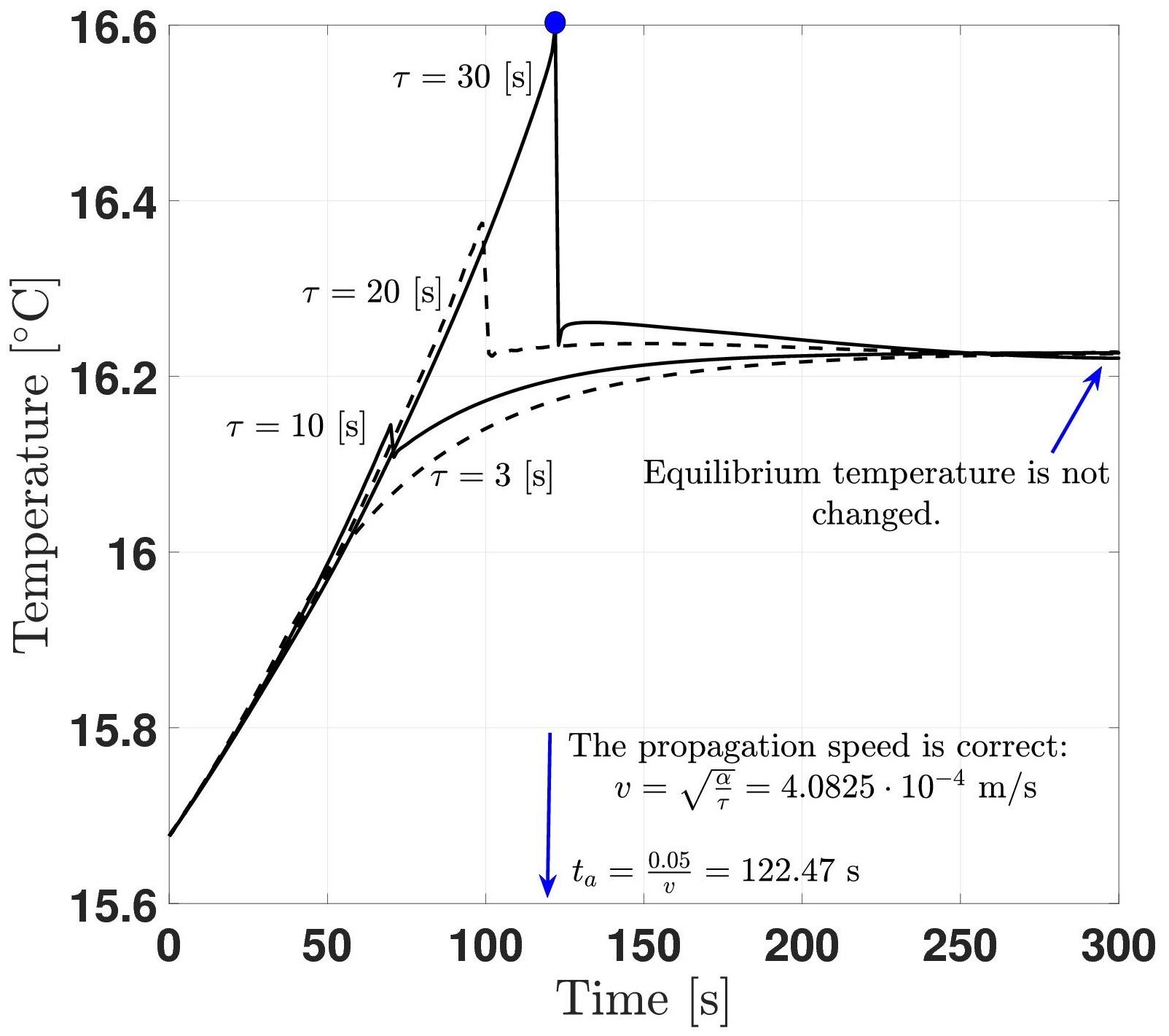}
\caption{Temperature history according to the MCV equation, at $x=0.05$ m, for different relaxation times ($\tau=\{0, 0.025, 0.05, 0.075, 0.1\}$ s), using $N=500$ terms.}
\label{fig4}
\end{figure}

In regard to the initial conditions, we aim to investigate the role of the $l^2$ term. Apparently, substituting the intial condition \eqref{icbc} into \eqref{gk0} and solving it as a partial differential equation, would offer us no advantage, similarly to \eqref{mcvic0}. Instead, we remain with the Galerkin method, using the same basis functions for $\phi(x)$ and $\varphi(x)$ in \eqref{g1}, thus we obtain
\begin{align}
\tau \frac{\textrm{d} a_j}{\textrm{d} t} + \Big (1 + l^2 \frac{j^2 \pi^2}{L^2} \Big ) a_j &= \lambda \frac{j \pi}{L} b_j, \label{g3gk}\\
\rho c \frac{\textrm{d} b_j}{\textrm{d} t}=& -\frac{j \pi}{L} a_j.\label{g3e}
\end{align}
Interestingly, that spatial derivative term of $l^2$ appears only in a coefficient of $a_j$, therefore it does not differ from the MCV equation significantly. In other words, the term $l^2$ does not provide further restrictions on the initial conditions. Furthermore, the same series expansion \eqref{f1} can be applied here as well. 
We find the temperature history in the form of
\begin{align}
b_j(t)=\frac{b_{j0}}{2 \beta_j} e^{-\frac{ (\beta_j+\gamma_j)t}{2 \tau }} \left[(\gamma_j-2 \alpha_j  \tau) \left(e^{\frac{ \beta_j t}{\tau }}-1\right)  +\beta_j \left(e^{\frac{\beta_j t}{\tau }}+1\right)\right]
\end{align}
with
\begin{align}
\alpha_j=\frac{\lambda}{\rho c} \frac{j^2 \pi^2 }{L^2}; \quad \beta_j=\sqrt{\gamma_j^2-4\alpha_j \tau}; \quad \gamma_j=1 + l^2 \frac{j^2 \pi^2}{L^2}.
\end{align}
Naturally, $l^2=0$ recovers the solution of the MCV equation. First, we illustrate this special case: Figure \ref{fig4} presents the characteristic solutions for the MCV equation with changing the relaxation time. As it is visible, in order to observe a wave propagation effect, extremely large relaxation time is needed. Since  the materials usually have much smaller relaxation time, it is not possible to experimentally observe such phenomenon under normal conditions on macroscale, in agreement with \cite{Aur16}.

\begin{figure}[H]
\centering
\includegraphics[width=12cm,height=7cm]{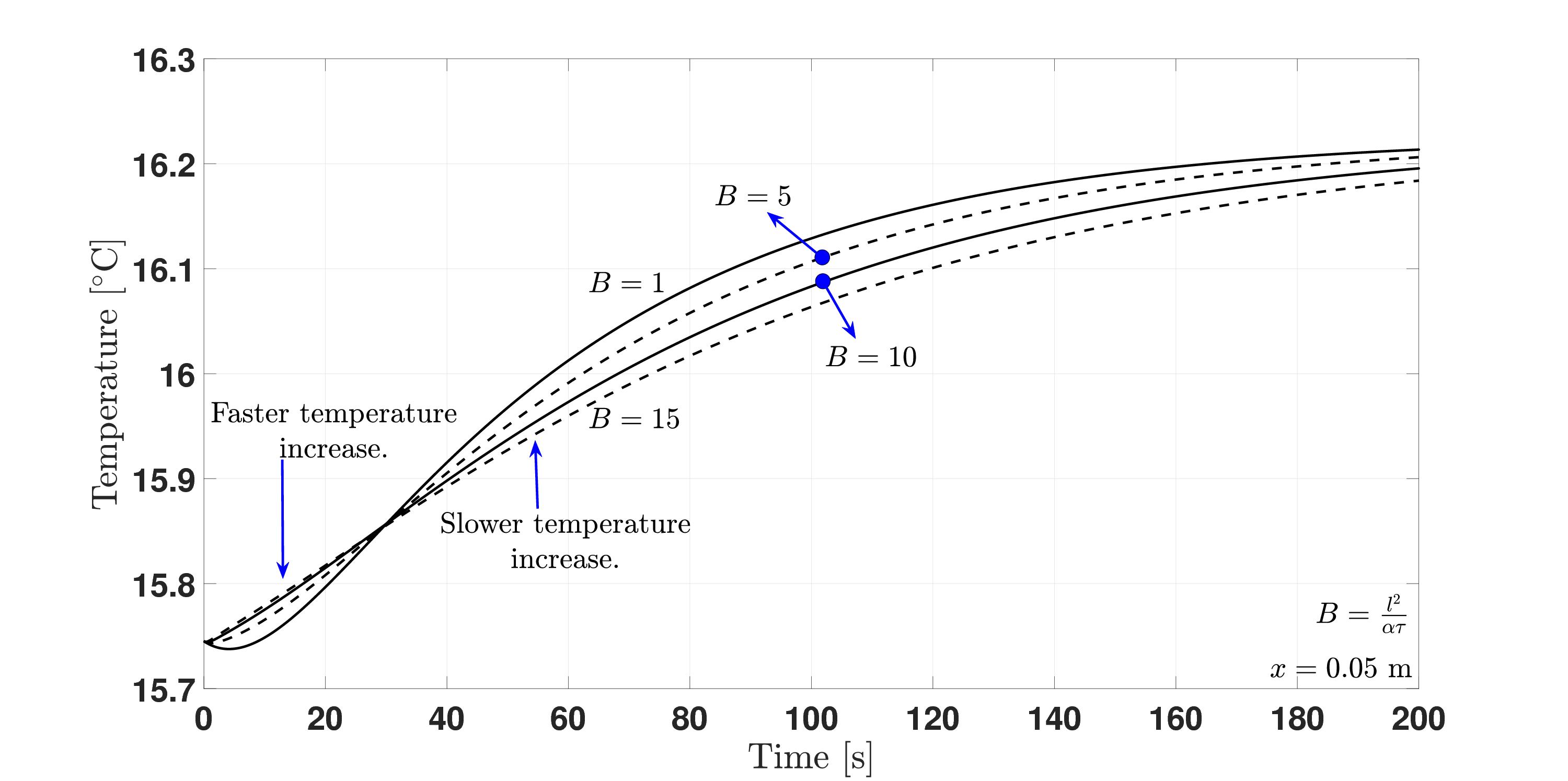}
\caption{Temperature history according to the GK equation, at $x=0.05$ m, for different  ($l^2=\{0.075, 0.375, 0.75, 1.125\}\cdot 10^{-4}$ m), using $N=5$ terms.}
\label{fig5}
\end{figure}

Next, Figure \ref{fig5} shows the behaviour of the GK equation, presenting the characteristic temperature curves at the middle. In this case, $l^2$ is changed in order to achieve the corresponding $B=l^2/(\alpha \tau)$ values. That $B$ is a dimensionless parameter, being helpful in the characterisation of the solution on how it deviates from the Fourier equation \cite{Vanetal17}. Fourier's solution is recovered when $B=1$ (`resonance' case, see \cite{VanKovFul15, FulEtal18e}). As $B$ increases, the temperature rises faster at the beginning, resulting in better convergence properties in agreement with \cite{FehKov21, Kov18gk}. After that initial time interval, the temperature rise becomes slower than the one belonging to the Fourier equation. This phenomenon is called over-diffusion and is characteristic to the GK equation only.

\section{Discussion of consequences}
A time will come shortly when the community must decide which non-Fourier equation should be the next standard model of heat conduction after Fourier's law. More and more possibilities are revealed, widening the application areas. Therefore, understanding these models, their physical consistency, and how to solve them have increasing importance for which this paper aimed to contribute.

For non-Fourier equations, the usual approach does not work as it used to do with the Fourier heat equation. The initial and boundary conditions must be handled with more care, as they do affect the overall outcome of the model. The solution methods must respect these attributes, even for the most straightforward linear situation. For nonlinear problems, such as temperature-dependent coefficients, the material parameters are functionally restricted as a consequence of the II.~law of thermodynamcis, one affecting the other, and without a consistent physical background, it is not possible to obtain a physically sound solution \cite{KovRog20}.

Throughout the paper, we showed that it is not advantageous to use the $T$-representation of the heat equations, as it hides the essential connections between the field variables and can influence how we think about the initial (and boundary) conditions. 

On the example of the MCV and GK equations, we presented the Galerkin-type solution method for nonhomogeneous initial condition, revealing that one needs to determine the heat flux first in order to be able to take the time derivative of the temperature at the initial state correctly into account. Even for a linear spatial dependence, the time derivative becomes non-zero, which impacts the entire solution. This is the main difference between Fourier's law and non-Fourier models: for the Fourier equation, it does not matter how far the system is from equilibrium, and it seems natural to consider it to be a static situation as it does not influence the time derivatives. However, if the system is initially in a non-equilibrium state, then it induces an initial non-zero time derivative for a non-Fourier equation. If this initial state is close to equilibrium, i.e., not far from a steady-state, then  Fourier's law can be applied to find the heat flux field at the initial time instant. For a numerical code, it can be interpreted as taking the $0^{th}$ step with the Fourier equation in determining the initial heat flux and then resuming the calculation with the non-Fourier model. 
On the contrary, if the initial state is far from equilibrium, it remains an open question how the heat flux field could be determined reliably.
Connected to this, a more general question also emerges: how can one
determine whether a state is close to or far from equilibrium? These
remain for further investigation.

\section{Acknowledgement}

The research reported in this paper and carried out at BME has been supported by the grants National Research, Development and Innovation Office-NKFIH FK 134277, by the NRDI Fund (TKP2020 NC, Grant No. BME-NC) based on the charter of bolster issued by the NRDI Office under the auspices of the Ministry for Innovation and Technology and the New National Excellence Program of the Ministry for Innovation and Technology project ÚNKP-21-5-BME-368. This paper was supported by the János Bolyai Research Scholarship of the Hungarian Academy of Sciences.

\bibliographystyle{unsrt}

\end{document}